\documentstyle[12pt]{article}
\renewcommand{\title}[1]{{\Large\bf\mbox{}\\\medskip#1\bigskip\medskip\\}}
\renewcommand{\author}[1]{{\large #1\smallskip\\}}
\newcommand{\address}[1]{{\em #1\medskip\\}}

\begin{document}
\begin{center}
\title{The Baxter Revolution}
\author{Barry M. McCoy}
\address{C. N. Yang Institute for Theoretical Physics\\
State University of New York\\
Stony Brook, NY 11794-3840, USA}
\begin{abstract}
\noindent

I review the revolutionary impact Rodney Baxter has had on statistical
mechanics beginning with his solution of the 8 vertex model in 1971
and the invention of corner transfer matrices in 1976 to the creation
of the RSOS models in 1984 and his continuing current work on the
chiral Potts model.

\end{abstract}
\end{center}

\section{Introduction}

At the beginning of the $20^{th}$ century statistical mechanics was
conceived of as a microscopic way to understand the laws of
thermodynamics and the kinetic theory of gases. In practice its scope
was limited to the classical ideal gas, 
the perfect quantum gases and finally to
 a diagrammatic technique devised in the $30's$ for computing the low density
properties of gases. At that time there was even debate as to if
the theory were in principle powerful enough to include phase
transitions and dense liquids.

All of this changed in 1944 when Onsager \cite{ons} demonstrated that
exact solutions of strongly interacting problems were possible 
by computing the free energy of the Ising model.
But, while of the greatest
importance in principle, this discovery did not radically alter the field of
statistical mechanics in practice. However, starting 
with the beginning of the 70's
Rodney Baxter took up the cause of exactly solvable models in
statistical mechanics and from that time on the field has been so
totally transformed that it may truly be said that a revolution has
occurred. In this paper I will examine how this revolution came about.

\section{The Eight Vertex model}

Onsager's work of 1944 was monumental but cannot be said to be
revolutionary because its consequences were so extremely
limited. Kaufman and Onsager \cite{kauf}--\cite{ko} 
reduced the computations to a free  fermi
problem in 1949 and after Yang \cite{yang} computed 
the spontaneous magnetization
in 1952 there were no further developments. Indeed the reduction of
the solution of the Ising model to a free fermi problem had the effect
of suggesting that Onsager's techniques were so specialized that there
might in fact not be any other statistical mechanical models which
could be exactly solved.

It was therefore very important when in 1967 Lieb \cite{lieb}
introduced and solved (cases of) the six vertex model.
This showed that other exactly solvable
statistical mechanical problems did indeed exist.
Lieb found that
this statistical model had the very curious property that the
eigenvectors of its transfer matrix were exactly the same as the
eigenvectors of the quantum spin 1/2 anisotropic Heisenberg chain 
\begin{equation}
H=-{1\over 2}\sum_{j=1}^L(\sigma^x_j\sigma^x_{j+1}+
\sigma^y_j\sigma^y_{j+1}+\Delta \sigma^z_j\sigma^z_{j+1})
\label{hxxz}
\end{equation}
which had been previously solved \cite{orb}-\cite{yy} by methods 
that went back to the work
of Bethe \cite{bethe} in 1931.
This result is particularly striking because the six vertex model
depends on one more parameter that does the $XXZ$ spin chain. That
extra parameter (which I will refer to as $v$) appears in the
eigenvalues of the transfer matrix but not in the eigenvectors. 
The reasons for this curious relation
between the quantum spin chain in one dimension and the problem in
classical statistical mechanics in two dimensions were totally obscure.

At that time the author was a post doctoral fellow and he and his
thesis advisor in a completely obscure paper \cite{mw} 
explained the relation  between the quantum and classical system by
demonstrating that  the transfer matrix for the six vertex model $T(v)$
commutes for all $v$ with the Hamiltonian (\ref{hxxz})  of the XXZ model.
\begin{equation}
[T(v),H]=0.
\label{thcomm}
\end{equation} 
This commutation relation guarantees that the eigenvectors of $T(v)$
are independent of $v$ and that they are equal to the eigenvectors of
$H$ without having to explicitly compute the eigenvectors themselves.  

The next year Sutherland \cite{suth} found 
an identical commutation relation
between the quantum Hamiltonian of the XYZ model 
\begin{equation}
H_{\rm XYZ}=\sum_{j=1}^L(J^x\sigma^x_j \sigma^x_{j+1}+
J^y\sigma^y_j \sigma^y_{j+1}+ J^z\sigma^z_j \sigma^z_{j+1})
\end{equation}
and the transfer matrix of the eight-vertex model.
But since neither the eight vertex model nor the
$XYZ$ model had been solved this  commutation relation
merely related two equally intractable problems.

All mysteries were resolved when in 1971 Baxter solved both the
eight vertex model and the $XYZ$ model \cite{bax1}-\cite{baxpaper} 
at the same time and moreover
solved them by inventing methods of such power and generality that the
the course of research in 
statistical mechanics was permanently altered. This is the beginning
of the Baxter revolution.

The first revolutionary advance made by Baxter 
\cite{bax1}-\cite{baxpaper} was the generalization of
\begin{equation}
[T(v),H]=0 ~~{\rm to}~~[T(v),T(v')]=0
\label{commtran}
\end{equation}
and that
as $v\sim 0$
\begin{equation}
T(v)\sim T(0)(1+cv +vH_{XYZ})
\end{equation}
This generalization is of great importance because it
relates a model to itself and can be taken as a general criteria
which selects out particular models of interest. 
Moreover, Baxter demonstrated
the existence of this global commutation relation by means of a local
relation between Boltzmann weights. Baxter called this local relation 
a star triangle equation because the first such relation had already
been found by Onsager \cite{ons}, \cite{wan} in the Ising model and Onsager had referred to
the relation as a star triangle equation. A related local equation had
been known since the work of McGuire \cite{jim} and Yang \cite{yang3} on the
quantum delta function gases but its deep connection with the work of
Onsager had not been understood. 
The search for solutions of the star triangle equation has been of
major interest ever since and has led to the creation of the entirely
new field of mathematics called ``Quantum Groups'' \cite{drin}-\cite{jimbo}.  
The Baxter revolution of 1971 is directly
responsible for this new field of mathematics

The second revolutionary step in Baxter's paper \cite{bax1} is that
in addition to the commutation relation (\ref{commtran}) he was able to
obtain a functional equation for the eigenvalues of the transfer
matrix and from this he could obtain equations which characterized the
eigenvalues. In the limit where the eight vertex model becomes the six
vertex model these equations reduced to the Bethe's equations
previously found by Lieb \cite{lieb}. But Lieb found his equations by finding
expressions for all of the eigenvectors of the problem whereas Baxter
never considered eigenvectors at all. It is truly a revolutionary
change in point of view to divorce the solution of the eigenvalue and
eigenvector problems and to solve the former without knowing anything
of the latter. This technique has proven to be of  
utmost generality and, indeed, for almost every solution which has
been found to the star triangle equation a corresponding functional
equation for eigenvalues has been found. On the other hand, the study
of the eigenvectors, which was the heart of the solution of the six
vertex and XXZ models has almost been abandoned.

The final technique introduced by Baxter is the thorough going use of
elliptic functions. Elliptic functions, of course,  have been used in
physics since the days of the heavy symmetric top and are
conspicuously used in Onsager's solution of the Ising model.
But even though elliptic functions appear in Onsager's final
expression for the free energy of the Ising model they play no role 
in either Onsager's original algebraic solution or in Kaufman's free
fermi solution. On the other hand there are steps in Baxter's solution
where the elliptic functions are essential. It is quite fair to say
that just as Onsager invented  the loop group of $sl_2$ in his solution of the
Ising model that Baxter in his 1971 paper first introduced the
essential use of elliptic and modular functions into 20th century physics.

\section{The corner transfer matrix}

It took Onsager 5 years from the computation of the Ising model free
energy before he made public his conjecture for the order parameter
\cite{ons2}. Baxter was much more prompt in the case of the eight
vertex model and produced with Barber in 1973 
conjecture for the order parameter
\cite{barbax} a mere two years after the free energy was computed. For
the Ising model it took another three years to go from the conjecture
to a proof \cite{yang}. For the eight vertex model it also took Baxter
three years to obtain a proof of the conjecture.

The details of Baxter's proof are contained in two separate papers
 \cite{bax3}-\cite{bax4} and
form the subject of chapter 13 of his 1982 book {\it Exactly Solved
Models in Statistical Mechanics} \cite{bax5}. 
It is even more revolutionary than
the 1971 free energy computation. Baxter not only
abandons the use of the eigenvectors of the row to row transfer matrix
(which had been retained in his 1973 computation of the free energy of
the six vertex model order parameter \cite{bax6}) 
but he abandons the use of the
row to row transfer matrix altogether. In its place he uses a
completely new construct which had never been seen before and which
had absolutely no precursors in the literature: the corner transfer matrix.

A transfer matrix builds up a large lattice one row at a time. In an
$L \times L$ lattice of a 2 state per site model it has dimension $2^L.$
A corner transfer matrix builds up a lattice by adding one quadrant at
a time and has dimension $2^{L^2/4}.$ The spin whose average is being
computed lies at the corner common to all four quadrants.
Order parameters are computed from the eigenvector of the ground state 
of the row to row transfer matrix. For the corner transfer matrix the
order parameter is expressed in terms of the eigenvalues and the
eigenvectors are not needed.  

Thus far the philosophy of the order parameter computation has followed
the spirit of the free energy computation in that all attention has
been moved from eigenvectors to eigenvalues. But in order to make this
a useful tool Baxter takes one more revolutionary step. He takes the
thermodynamic limit before he obtains equations for the
eigenvalues. This is exactly the opposite from what was done in the
free energy computation where the equations are obtained first and only
in the end is the thermodynamic limit taken.

This early introduction of the thermodynamic limit has a very dramatic
impact on the eigenvalues of the corner transfer matrix. To see this
we note that the matrix elements of the corner transfer matrix are all
quasi periodic functions of the spectral variable $v.$ This is of
course also true for the row to row transfer matrix. It is thus a
natural argument to make to say that a matrix
with quasi periodic elements should be have doubly periodic
eigenvalues and this is in fact true for the row to row transfer
matrix. But for the corner transfer matrix the taking of the
thermodynamic limit has the astounding effect that the  eigenvalues,
instead of being elliptic functions all become simple exponentials
$e^{-\alpha_r v}.$ Once these very simple exponential expressions for
the eigenvalues are obtained it is a straightforward matter to obtain the final
form for the spontaneous magnetization of the eight  vertex model, but
all along the way, it is fair to say, a great deal of magic has been worked.

\section{The RSOS models}

The next stage in the Baxter revolution is the discovery and solution
of the RSOS model by Andrews, Baxter and Forrester in 1984 \cite{abf}.
As in the case of the eight vertex model revolution in 1971   there were
several precursor papers, this time all by Baxter himself.

It has been stressed in the preceding sections that Baxter made a
revolutionary shift of point of view by discovering that the
eigenvalue problems could be solved without solving the eigenvector problems. 
Therefore for the six vertex and $XXZ$ model Baxter could obtain the Bethe's
equations for the eigenvalues without recourse to the Bethe's form of
the eigenvector
\begin{equation}
\psi(x_1,x_2,\cdots, x_n)=\sum_P A(P)e^{i\sum_j x_j k_{Pj}}.
\label{beqn}
\end{equation}
In the previous work on the six vertex and $XXZ$ models the
restriction was made that all the $k_j$ were distinct. It was
therefore quite a surprise when in 1973 Baxter discovered \cite{bax7} 
that in the $XXZ$ chain (\ref{hxxz}) when
\begin{equation}
\Delta={1\over 2}(q+q^{-1})  ~~~{\rm and}~~q^{2N}=1
\label{unity}
\end{equation}
that there are in fact eigenvectors of the $XXZ$ chain for which the
$k_j$ of (\ref{beqn}) are equal. For these solutions the $k_j$ obey
\begin{equation}
\Delta e^{ik_j}-1-e^{i(k_j+k_l)}=0
\label{rest}
\end{equation}
and this case had been tacitly excluded in all previous work.

In \cite{bax7} Baxter generalized the root of unity
condition  (\ref{unity}) of the six vertex model to the eight vertex
model and he found an entire basis of eigenvectors which in a sense
makes maximal use of the violation of the previously assumed
condition $k_j\neq k_l$. Baxter is thus able to re-express these root
of unity eight vertex models in terms of what he calls in his 1973
paper an ``Ising--like model.''

Baxter's next encounter with root of unity models was in 1981 when he
solved the hard hexagon model \cite{bax8}. In this most remarkable
paper Baxter uses his corner transfer matrices to compute the order parameter 
of the problem and in the course of the computation discovers the
identities of Rogers \cite{rog} and Ramanujan \cite{rr} which were
first found in 1894
\begin{equation}
\sum_{n=0} {q^{n(n+a)}\over (q)_n}={1\over
(q)_{\infty}}\sum_{n=-\infty}^{\infty}(q^{n(10n+1+2a)}-q^{(5n+2-a)(2n+1)})
\label{rriden}
\end{equation}
where $(q)_n=\prod_{j=1}^n(1-q^j)$ and $a=0,1.$
Baxter was clearly impressed that these classic identities appeared
naturally in a statistical mechanics problem because he put the term
``Rogers--Ramanujan'' in the title of the paper. Because the right hand
side of (\ref{rriden}) is obviously written as the difference of two
theta functions we once again see that modular functions appear
naturally in statistical mechanics.
But neither the 1973 nor the 1981 papers can be called genuinely revolutionary 
because neither of them was seen to have general applicability.

The revolution that allowed the general applicability of Baxter's
techniques is carried out in the paper of 1984 with Andrews and
Forrester \cite{abf} and the companion paper by Forrester and Baxter
\cite{fb} in which it was shown that the hard hexagon model of
\cite{bax8} is obtained from a special case of 
the ``Ising--like models'' found in
the root of unity  eight vertex models in 1973 \cite{bax7}. These
Ising--like models are now called eight vertex solid--on--solid models 
and the restriction needed to obtain the hard hexagon model is in the
general case called the restricted solid--on--solid model.
Starting from this formulation of the RSOS models the order parameters
are computed by a direct application of the corner transfer matrix
method and at the step where in the hard hexagon model the identity 
(\ref{rriden}) was obtained the authors of \cite{abf},\cite{fb}
instead solve a path counting problem and find the general result
 \begin{equation}
{1\over (q)_{\infty}}\sum_{n=-\infty}^{\infty}(q^{n(npp'+rp'-sp)}-
q^{(np'+s)(jp+r)})
\label{roc}
\end{equation}
where the relatively prime integers $p$ and $p'$ effectively
parameterize the root of unity condition (\ref{unity}).
This sum in this result is obviously the difference of two Jacobi
theta functions and thus we see that all the RSOS models lead to theta 
functions. But most remarkably the exact same expression (\ref{roc})
was discovered at the same time to arise in the expression of the
characters \cite{ff}-\cite{rocca} of the minimal 
models $M(p,p')$ conformal field theory \cite{bpz} and
these models were soon thereafter obtained as cosets \cite{gko} 
of the affine Lie algebra $A_1^{(1)}.$  

It thus became clear that the statistical
mechanics of RSOS models, conformal field theory, and affine Lie
algebras are all part of the same subject and from this point forth
the results of statistical mechanics appear in such apparently
unrelated fields as string theory ,number theory and knot theory. 
Baxter's corner
transfer matrix was seen to be intimately related to constructions in
the theory of affine Lie algebras involving null vectors and  the
corner transfer matrix computations of Baxter's
statistical models were rapidly generalized from the affine Lie algebra
 $A_1^{(1)}$ to all affine Lie algebras. Solvable statistical mechanical
models were now seen everywhere in physics and Baxter's methods were
subject to vast generalization.

\section{ The chiral Potts model}

For a few years it was thought that the revolution was complete
and that corner transfer matrix methods and group theory could solve
all problems which started out from commuting transfer matrices. This
was changed however when the chiral Potts model was discovered in 1987
\cite{ampty}. This model does indeed satisfy the condition of
commuting transfer matrices (\ref{commtran}) and the Boltzmann
weights do  
obey a star triangle equation but unlike all previously seen models
the Boltzmann weights are not parameterized either by trigonometric or
elliptic functions but rather are functions on some higher genus
spectral curve. There is a modulus like 
variable $k$ in the model and when $N=3$
the genus of the curve is 10 if $k\neq 0,1$ and if $k=1$ the curve is
the very symmetric elliptic curve $x^3+y^3=z^3.$
If $N=4$ and $k=1$ the curve is the fourth order Fermat curve
$x^4+y^4=z^4$ which has genus three \cite{sah}.
 
As would be expected Baxter rapidly became interested in this problem  
and soon Baxter. Perk and Au--Yang \cite{bpa} found that for arbitrary
$N$ and $k$ the spectral
curve has the very simple form
\begin{equation}
a^N+kb^N=k'd^N~~~{\rm and}~~~ka^N+b^N=k'c^N
\label{cp}
\end{equation}
with $k^2+k'^2=1.$
When $N=2$ this curve reduces to an elliptic curve and the chiral
Potts model reduces to the Ising model. However, in general for $k\neq
0,1$ the curve has genus $N^3-2N^2+1$ and for $k=1$ the curve reduces
to the $N^{th}$ order Fermat curve of genus $(N-1)(N-2)/2.$

The first thing to attempt after finding the Boltzmann weights for
the chiral Potts model
is to repeat what had been done so many times before and to obtain a
functional equation for the eigenvalues . 
That was soon done \cite{bax9}-\cite{bax10} but the
next step in Baxter's program was not so easy because the methods of
solution of this functional equation which relied on the properties of
genus 1 elliptic functions did not work. Solutions for the free energy
which by passed the elliptic functions were soon found
\cite{bax9}--\cite{bax11} but the fact that new methods were needed
indicated that 
the revolution was not yet complete.

The greatest puzzle was set up in 1989 when after generalizing
earlier work on the $N=3$ state model \cite{kad} it was conjectured \cite{amp}
on the basis of extensive series expansions 
that the order parameters of the $N$ state chiral Potts model are
given by
\begin{equation}
M_n=(1-k^2)^{n(N-n)/2N^2} ~~~{\rm for}~~1\leq n \leq N-1
\label{cporder}
\end{equation}
This remarkably simple expression reduces to the result of Onsager 
\cite{ons2} and Yang \cite{yang} for the Ising model when $N=2$ and is
a great deal simpler than the order parameters for the RSOS models
\cite{abf}, \cite{fb}.
The first expectation was that Baxter's corner transfer matrix methods
could be applied to prove the conjecture true and the first attempt to
do this was made by Baxter in \cite{bax12}. 
In this paper Baxter gives a new and very transparent derivation of
the corner transfer matrix methods and he reduces the computation of
the order parameter to a problem of the evaluation of a path ordered
exponential of non--commuting operators over a Riemann surface. Such a
formulation sounds as if methods of non Abelian field theory could
now be applied to solve the problem. Unfortunately 
to quote Baxter in
a subsequent paper \cite{bax13} ``Surprisingly the method completely
fails for the chiral Potts model.''

The reason for the failure of the method is that the introduction of
the higher genus curve into the problem has destroyed a property used
by Baxter and all subsequent authors in the application of corner
transfer matrix methods. This property is the so called difference
property which is the property, shared by the plane and torus but by
no curve of higher genus, of having an infinite automorphism group
(the translations). It is this property which was used to reduce the
eigenvalues to exponentials in the spectral variable and it is not
present in the chiral Potts model. 

\section{Future Prospects}

The discovery of the chiral Potts model has made it now clear that the
Baxter revolution has met up with  problems in algebraic geometry
which have proven intractable for almost 150 years. Baxter has
investigated these problems now for almost a decade
\cite{bax12}--\cite{bax16}
 and it is clear
that the solution of these physics problems will make a major advance
in mathematics. But even with this evaluation of current problems the
impact of Baxter's revolution is clearly seen. Mathematics is no
longer treated as a closed finished subject by physicists. More than
anyone else Baxter has taught us that physics guides mathematics and
not the other way around. This is of course the way things were in the
$17^{th}$ century when Newton and Leibnitz invented calculus to study
mechanics. Perhaps in the intervening centuries in the name of being
experimental scientists we physicists drifted away from 
away from doing creative mathematics. The work of Rodney Baxter
serves now and will serve in the future as a beacon of inspiration
to all those who believe  that there is a unity in physics and
mathematics  which provides inspiration that can be obtained in no
other way.

\bigskip
{\bf Acknowledgments}

This work was partially supported by NSF grant DMR97--03543.


\begin{thebibliography}{99}
\bibitem{ons} L. Onsager, Phys. Rev. 65 (1944) 117.



\bibitem{kauf}B. Kaufman, Phys. Rev. 76 (1949), 1232.
\bibitem{ko} B. Kaufman and L. Onsager, Phys. Rev. 76 (1949) 1244.
\bibitem{yang} C.N. Yang, Phys. Rev. 85 (1952) 808.
\bibitem{lieb}E. Lieb, Phys. Rev. Letts. 18 (1967) 692; 18 (1967)
1046; 19 (1967) 108; Phys. Rev. 162 (1967) 162.
\bibitem{orb}R.Orbach, Phys. Rev. 112 (1958) 309.
\bibitem{walker} L.R. Walker, Phys. Rev. 116 (1959) 1089.
\bibitem{gaud}J. des Cloizeaux and M. Gaudin, J. Math. Phys. 7 (1966) 1384. 
\bibitem{yy} C.N. Yang and C.P. Yang, Phys. Rev. 150 (1966) 321; 327 (1966)
\bibitem{bethe} H.A. Bethe, Z. Physik 71 (1931) 205.
\bibitem{mw} B.M. McCoy and T.T. Wu, Il Nuovo Cimento, 56 (1968) 311.
\bibitem{suth} B. Sutherland, J. Math. Phys. 11 (1970) 3183.
\bibitem{bax1} R.J. Baxter, Phys.Rev. Lett. 26 (1971) 832.
\bibitem{bax2} R.J. Baxter, Phys. Rev. Lett. 26 (1971) 834.
\bibitem{baxpaper} R.J.Baxter, Ann. Phys. 70 (1972) 193, 323.
\bibitem{wan} G.H. Wannier, Rev. Mod. Phys. 17 (1945) 50.
\bibitem{jim}J.B. McGuire, J. Math. Phys. 5 (1964) 622; 6 (1965) 432;7
(1967) 123.
\bibitem{yang3}C.N. Yang, Phys. Rev. Lett. 19 (1967) 1312.
\bibitem{drin} V.G. Drinfel'd, Soviet Math. Doklady 32 (1985) 254.
\bibitem{jimbo}M. Jimbo, Lett. Math. Phys. 10 (1985) 63.
\bibitem{ons2}L. Onsager , discussion, Nuovo Cimento 6, suppl., 261
(1949).
\bibitem{barbax}M.N. Barber and R.J. Baxter, J. Phys. C 6 (1973) 2913.
\bibitem{bax3} R.J. Baxter, J. Stat. Phys. 15 (1976) 485.
\bibitem{bax4} R.J. Baxter, J. Stat. Phys. 17 (1977) 1.
\bibitem{bax5}R.J. Baxter, {\it Exactly Solved Models in Statistical
Mechanics}, (Academic Press 1982).
\bibitem{bax6}R.J. Baxter, J. Stat. Phys. 9 (1973) 145.
\bibitem{abf} G.E. Andrews, R.J. Baxter and P.J. Forrester,
J. Stat. Phys. 35 (1984) 193.
\bibitem{bax7}R.J. Baxter, Ann. Phys. 76 (1973) 1,14,48. 
\bibitem{bax8} R.J. Baxter, J. Stat. Phys. 26 (1981) 427.
\bibitem{rog}L.J. Rogers, Proc. Lond. Math. Soc. 25 (1894) 318.
\bibitem{rr}L.J. Rogers and S. Ramanujan, Proc. Camb. Phil. Soc. 19
(1919) 315.

\bibitem{fb} P.J. Forrester and R.J. Baxter, J. Stat. Phys. 38 (1985) 435.
\bibitem{ff}B.L. Feigin and D.B. Fuchs, Funct. Anal. and Appl. 16
(1982) 114.
\bibitem{rocca} A. Rocha--Caridi, in {\it Vertex Operators in
Mathematics and Physics} Eds. J. Lepowsky, S. Mandelstam and
I. Singer, Springer--Verlag, New York (1985) 451.
\bibitem{bpz}A.A. Belavin, A.M. Polyakov and A.B. Zamolodchikov,
J. Stat. Phys. 34 (1984) 763; Nucl. Phys. B241 (1984) 333.
\bibitem{gko}P. Goddard, A. Kent and D. Olive, Phys. Lett. B152 (1985)
88; Comm. Math. Phys. 103 (1986) 105.

\bibitem{ampty}H.Au--Yang, B.M. McCoy, J.H.H. Perk, S. Tang and M--L
Yan, Phys. Lett. A123 (1987) 219.
\bibitem{sah} B.M. McCoy, J.H.H. Perk, S. Tang and C.H. Sah,
Phys. Lett. A 125 (1987) 9.
\bibitem{bpa}R.J. Baxter, J.H.H. Perk and H. Au--Yang. Phys. Letts
A128 (1988) 138.
\bibitem{bax9} R.J. Baxter, Phys. Lett. A133 (1988) 185.
\bibitem{mccoy1} G. Albertini, B.M. McCoy and J.H.H. Perk,
Phys. Lett. A135 (1989) 159.
\bibitem{mccoy2} G. Albertini, B.M. McCoy and J.H.H. Perk,
Phys. Lett. A 139 (1989) 204; and in {\it Advanced Studies in Pure
Mathematics}, Vol. 19 ed. M. Jimbo, T. Miwa and A. Tsuchiya,
(Kinokuniya--Academic Press, Tokyo 1989) 1.
\bibitem{bazstr} V.V. Bazhanov and Yu. G. Stroganov, J. Stat. Phys. 51
(1990) 799.
\bibitem{bax10} R.J. Baxter, V.V. Bazhanov and J.H.H. Perk,
Int. J. Mod. Phys. 4 (1990) 803.
\bibitem{bax11}R.J. Baxter, J. Stat. Phys. 52 (1988) 639.
\bibitem{kad} S. Howes, L.P. Kadanoff and M. den Nijs,
Nucl. Phys. B215[FS7] (1983) 169.
\bibitem{amp}G. Albertini, B.M. McCoy, J.H.H. Perk, and S. Tang,
Nucl. Phys. B314 (1989) 741.
\bibitem{bax12} R.J. Baxter, J. Stat. Phys. 70 (1993) 535.
\bibitem{bax13} R.J. Baxter, J. Stat. Phys. 91 (1998) 499.
\bibitem{bax14} R.J. Baxter in {\it Proceedings of the International
Congress of Mathematicians} (Springer Verlag 1990) 1309.
\bibitem{bax15}R.J. Baxter, J. Phys. A 31 (1998) 6807.
\bibitem{bax16} R.J. Baxter, Physica A 260 (1998) 117.

\end{thebibliography}
\end{document}